\documentclass[a4paper,10pt]{article}
\usepackage{amsfonts}
\usepackage{amsmath}
\usepackage{amsthm}
\usepackage{amssymb}
\usepackage{amsmath, graphics}
\usepackage{epsf,amsfonts,amsmath}
\newcommand{\mathsym}[1]{{}}

\begin{document}
\title{\textbf{Mayer Transfer Operator Approach to Selberg Zeta Function}}
\author{\begin{Large}Arash Momeni$^1$ and Alexei Venkov$^2$\end{Large}\\ \begin{small}
$^1$ Department of Statistical Physics and Nonlinear Dynamics\end{small}\\ \begin{small}Institute of Theoretical Physics, \end{small}\\
\begin{small}Clausthal University of Technology, 38678 Clausthal-Zellerfeld, \end{small}\\ \begin{small}Germany. E-Mail: arash.momeni@tu-clausthal.de \end{small}\\
\begin{small}$^2$Institute for Mathematics and Centre of Quantum Geometry QGM,\end{small}\\ \begin{small}University of Aarhus, 8000 Aarhus C,\end{small}\\ \begin{small}Denmark. E-Mail: venkov@imf.au.dk \end{small}\\}
\maketitle

\begin{abstract}
These notes are based on three lectures given by the second author at Copenhagen University (October
2009) and at Aarhus University, Denmark (December 2009). We mostly present here a survey of results of Dieter Mayer on relations between Selberg and Smale-Ruelle dynamical zeta functions. In a special situation the dynamical zeta function is defined for a geodesic flow on a hyperbolic plane quotient by an arithmetic cofinite discrete group. More precisely, the flow is defined for the corresponding unit tangent bundle. It turns out that the Selberg zeta function for this group can be expressed in terms of a Fredholm determinant of a classical transfer operator of the flow. The transfer operator is defined in a certain space of holomorphic functions and its matrix representation in a natural basis is given in terms of the Riemann zeta function and the Euler gamma function.
\end{abstract}

\section{General theory}
We quote Ruelle \cite{Ru,Ru1} to introduce his general notion of a transfer operator and a dynamical zeta function for a given dynamical system. 

First we give the definition of a weighted dynamical system. Let $\Lambda$ be a set weighted by a function $g:\Lambda\longrightarrow \mathbb C$. Assume that $\Lambda$ describes a system, then the dynamics of the system is given by a map $F:\Lambda\longrightarrow\Lambda$. The triplet $\mathcal D:=(\Lambda, F, g)$ is called a weighted dynamical system or simply a dynamical system. 

The transfer operator method is applicable if the map $F$ is not invertible, that is, for example, when its inverse is not unique. More precisely the set of inverse branches of $F$ must be finite or at least countable and discrete in a natural topology. 

For such a dynamical system the Ruelle transfer operator $\mathcal L$ which acts on a function $f:\Lambda\longrightarrow\mathbb C$ is defined by  
\begin{equation}\label{tran-op-gen}
(\mathcal Lf)(x)=\sum_{y\in F^{-1}\left\lbrace x\right\rbrace }g(y)f(y)
\end{equation}
One can see that the set of transfer operators for all dynamical systems of the set $\Lambda$ with respect to the product $\circ$ given by $(\mathcal L_1\circ\mathcal L_2)f=\mathcal L_1(\mathcal L_2f)$ becomes an algebra denoted by $\mathcal S$. A trace on this algebra is a linear functional $Tr:S\longrightarrow \mathbb C$ such that $Tr(\mathcal L_1\mathcal L_2)=Tr(\mathcal L_2\mathcal L_1)$ for every $\mathcal L_1$ and $\mathcal L_2$ in $\mathcal S$. For a given trace $Tr$ one can define formally a determinant $Det$ for the operators of the algebra by
\begin{equation}\label{det}
Det(I-z\mathcal L)=exp(-\sum_{m=1}^{\infty}\dfrac{z^m}{m}Tr\mathcal L^m)
\end{equation}

On the other hand a weighted dynamical system $\mathcal D=(\Lambda,F,g)$ is equipped with the so called Ruelle dynamical zeta function defined by
\begin{equation}\label{gen-z}
\zeta(z)=exp(\sum_{m=1}^\infty\dfrac{z^m}{m}\sum_{x\in FixF^m}\prod_{k=0}^{m-1}g(F^kx))
\end{equation}
where $FixF^m$ denotes the set of all fixed points of $F^m$. The set $FixF^m$ is finite or countably infinite for all $m>0$. Similar to other zeta functions the Ruelle dynamical zeta function has some sort of Euler product 
\begin{equation}\label{E-z}
\zeta(z)=\prod_{\left\lbrace P\right\rbrace }(1-z^{\vert P\vert}\prod_{k=0}^{\vert P\vert-1}g(F^kx_P))^{-1}
\end{equation}
where $\left\lbrace P\right\rbrace $ denotes the set of periodic orbits of $F$ with length $\vert P\vert$ and $x_P$ is an arbitrary element of $P$. We will assume that (\ref{det}), (\ref{gen-z}) and (\ref{E-z}) are absolutely convergent at least for $z$ from a certain domain in $\mathbb C$.

In general, analytic properties of zeta functions give an important information about the corresponding systems in question. For example a Tauberian theorem yields the classical prime number theorem from the positions of poles and zeros of the Riemann zeta function in the critical strip. 

In the same way we are interested in the analytic properties of the dynamical zeta function to get more information about the corresponding dynamical system. 

One of the important methods to study the analytic properties of dynamical zeta functions is the transfer operator method. In this method the analytic properties of the zeta function are related to the spectral properties of a transfer operator through a connection between the Fredholm determinant of the transfer operator and the dynamical zeta function.

One of the most important realization of the general programme described above is the Mayer transfer operator acting on some Banach space of holomorphic functions on a disc \cite{Mayer91}. This operator is assigned to the dynamical system related to the geodesic flow on the hyperbolic plane mod an arithmetical cofinite discrete group $\Gamma$. In this case the Fredholm determinant of the transfer operator is equal to the Selberg zeta function for the corresponding discrete group which is one of the most important aspects of Mayer's transfer operator theory. Indeed this equality provides us a new insight to the theory of quantum chaos. It turns out that the Mayer transfer operator which is a purely classical object surprisingly contains all information we can obtain from the corresponding Schroedinger operator.  
\section{Mayer's transfer operator for $PSL(2,\mathbb Z)$}
\theoremstyle{plain}\newtheorem{rem1}{Remark}
\theoremstyle{plain}\newtheorem{matrix-rep1}[rem1]{Remark}
\theoremstyle{plain}\newtheorem{rem2}[rem1]{Remark}
\theoremstyle{plain}\newtheorem{rem3}[rem1]{Remark}
\theoremstyle{plain}\newtheorem{lem1}{Lemma}
\theoremstyle{plain}\newtheorem{lem2}[lem1]{Lemma}
\theoremstyle{plain}\newtheorem{lem3}[lem1]{Lemma}
\theoremstyle{plain}\newtheorem{lem4}[lem1]{Lemma}
\theoremstyle{plain}\newtheorem{matrix-rep}[lem1]{Lemma}
\theoremstyle{plain}\newtheorem{lem5}[lem1]{Lemma}
\theoremstyle{plain}\newtheorem{lem0}[lem1]{Lemma}
\theoremstyle{plain}\newtheorem{lem6}[lem1]{Lemma}
\theoremstyle{plain}\newtheorem{lem7}[lem1]{Lemma}
\theoremstyle{plain}\newtheorem{lem8}[lem1]{Lemma}
\theoremstyle{plain}\newtheorem{lem9}[lem1]{Lemma}
\theoremstyle{plain}\newtheorem{lem10}[lem1]{Lemma}
\theoremstyle{plain}\newtheorem{lem11}[lem1]{Lemma}
\theoremstyle{plain}\newtheorem{LL'}[lem1]{Lemma}
\theoremstyle{plain}\newtheorem{defdd}{Deffinition}
\theoremstyle{plain}\newtheorem{cor1}{Corollary}
\theoremstyle{plain}\newtheorem{cor2}[cor1]{Corollary}
\theoremstyle{plain}\newtheorem{cor3}[cor1]{Corollary}
\theoremstyle{plain}\newtheorem{cor4}[cor1]{Corollary}
\theoremstyle{plain}\newtheorem{cor5}[cor1]{Corollary}
\theoremstyle{plain}\newtheorem{th1}{Theorem}
We start by introducing some notations and definitions. The hyperbolic plane $\mathbb H$ is the upper half plane $\left\lbrace x+iy\in\mathbb C \ \vert y>0\right\rbrace $ equipped with the Poincare metric $ds^2=y^{-2}(dx^2+dy^2)$ and the measure $d\mu (z)=y^{-2}dxdy$. Thus geodesics on $\mathbb H$ are the semicircles with centra and the end points on the real axis. 

The group of all orientation preserving isometries of the hyperbolic plane $\mathbb H$ is identified with the group 
\begin{eqnarray}
PSL(2,\mathbb R)=SL(2,\mathbb R)/\left\lbrace \pm I\right\rbrace 
\end{eqnarray}
acting on $\mathbb H$ by linear fractional transformations defined by  
\begin{displaymath}
z\rightarrow gz=\dfrac{az+b}{cz+d} \ \ \ \ \ g=\left( \begin{array}{cc}
a&b\\
c&d\\
\end{array}
\right)\ \ \ \ \ z\in \mathbb H.
\end{displaymath}

The modular group $PSL(2,\mathbb Z)$ is a discrete subgroup of $PSL(2,\mathbb R)$ defined by
\begin{equation}
PSL(2,\mathbb Z)=\left\lbrace \left( \begin{array}{cc}
a&b\\
c&d\\
\end{array}
\right)\vert \ ad-bc=1, \ a,b,c,d\in \mathbb Z\right\rbrace /\left\lbrace \pm I\right\rbrace.
\end{equation}
This is a non-cocompact Fuchsian group of the first kind. 

Let $M=PSL(2,\mathbb Z)\backslash \mathbb H$ be the quotient space of the hyperbolic plane $\mathbb H$ mod $PSL(2,\mathbb Z)$. This is a surface with one cusp and two conical singularities. Consider the continuous dynamics given by the geodesic flow $\varphi_t$ on $T_1M$, the unit tangent bundle of $M$. From the physics point of view, the unity of the tangent bundle means that the geodesic flow describes the motion of a free particle on $M$ with unit magnitude of velocity. 

As mentioned already, to define a transfer operator the corresponding dynamical map should have a finite or countable set of inverse branches while the geodesic flow is continuous and defines an invertible map on $T_1M$. Thus we first discretize the geodesic flow by constructing a Poincare map of $\varphi_t$. It is known that by a suitable choice of the Poincare section in $T_1M$ the dynamics of $\varphi_t$ reduces to the Poincare map given by (see \cite{chang})
\begin{eqnarray}
&P:[0,1]\times [0,1]\times\mathbb Z_2\rightarrow[0,1]\times [0,1]\times\mathbb Z_2&\nonumber\\&
(x_1,x_2,\epsilon)\mapsto(T_Gx_1,\dfrac{1}{[\frac{1}{x_1}]+x_2},-\epsilon)&
\end{eqnarray}
where $[x]$ denotes the integer part of $x$ and
\begin{equation}\label{Gauss-map}
T_Gx=\left\{ \begin{array}{ll}\frac{1}{x} \ mod\ 1&x\in(0,1]\\0&x=0
\end{array}\right.
\end{equation}
is the Gauss map.
\begin{rem1}
The group $PSL(2,\mathbb Z)$ does not contain the reflection relative to the y-axis. Consequently for every geodesic on $M$ there exists a unique geodesic on $M$ such that their representatives on the upper half plane are located symmetric relative to the $y$-axis. The same assertion holds for the orbits on $T_1M$. This fact is reflected by two possible values of the parameter $\epsilon$.
\end{rem1}
We are interested in the expanding part of the map $P$ which reflects the ergodic aspects of the geodesic flow $\varphi_t$, 
\begin{eqnarray}\label{dy-map-sl}
&P_{ex}:[0,1]\times\mathbb Z_2\rightarrow[0,1]\times\mathbb Z_2&\nonumber\\&
(x,\epsilon)\mapsto(T_Gx,-\epsilon).&
\end{eqnarray}

It remains to select a suitable weight function. Mayer chose the following weight function,
\begin{equation}\label{weight-fun}
g(x,\epsilon)=(T_G')^{s}(x)=x^{2s}
\end{equation}
where $s$ is a complex parameter. In fact, according to Sinai's work \cite{Sinai}, the ergodic properties of $\varphi_t$ are related to this weight function.

The Mayer transfer operator is a transfer operator for the weighted dynamical system \begin{equation}\label{dy-sy}
\mathcal D_1=([0,1]\times\mathbb Z_2,P_{ex},g(x,\epsilon)=x^{2s}),
\end{equation}
that is
\begin{equation}\label{mop}
\mathcal L_sf(x,\epsilon)=\sum_{y={P_{ex}^{-1}(x,\epsilon)}}g(y) f(y).
\end{equation}
We notice that for $s=1$ the Mayer transfer operator reduces to the Perron-Frobenious operator.
The map $P_{ex}$ has an infinite number of discrete inverse branches given by  
\begin{equation}\
P_{ex}^{-1}(x,\epsilon)=(\dfrac{1}{x+n},-\epsilon), \ \ \ n\in\mathbb N
\end{equation}
thus the Mayer transfer operator is formally expressed as 
\begin{equation}\label{Mayer1}
\overset{\sim}{\mathcal L}_sf(x,\epsilon)=\sum_{n=1}^\infty (\dfrac{1}{x+n})^{2s}f(\dfrac{1}{x+n},-\epsilon),\ \ \ \ \epsilon=\pm 1.
\end{equation}
Regarding the close connection of the weighted dynamical system (\ref{dy-sy}) to the group $PSL(2,\mathbb Z)$, this operator sometimes is called the Mayer transfer operator for $PSL(2,\mathbb Z)$. Equivalently we can write the Mayer operator (\ref{Mayer1}) in a vector form
\begin{equation}\label{Mayer vect}
\overset{\sim}{\mathcal L}_s\overrightarrow{f}(x)=\sum_{n=1}^\infty (\dfrac{1}{x+n})^{2s}M.\overrightarrow{f}(\dfrac{1}{x+n})
\end{equation}
with 
\begin{equation}
M=\left( \begin{array}{cc}
0&1\\
1&0\\
\end{array}
\right)\ \ \ \ \ \overrightarrow{f}(x)=\left( \begin{array}{c}
f(x,1)\\
f(x,-1)\\
\end{array}
\right)
\end{equation}

If instead of the group $PSL(2,\mathbb Z)$ we take the group \begin{equation}
PGL(2,\mathbb Z)=GL(2,\mathbb Z)/\left\lbrace \pm 1\right\rbrace
\end{equation}
 where

\begin{equation}
GL(2,\mathbb Z)=\left\lbrace \left( \begin{array}{cc}
a&b\\
c&d\\
\end{array}
\right)\vert \ ad-bc=\pm 1, \ a,b,c,d\in \mathbb Z\right\rbrace 
\end{equation}
then the orbits corresponding to the two possible values of $\epsilon$ will be identified, because $PGL(2,\mathbb Z)$ contains the reflection relative to the y-axis. Note that this reflection acts on $\mathbb H$ not by a linear fractional transformation but as the map $z\rightarrow-\overline{z}$. Therefore the dynamical system $\mathcal D_1$ in (\ref{dy-sy}) reduced to the following dynamical system,
\begin{equation}
\label{dy-sy2}
\mathcal D_2=([0,1], T_G, g(x)=x^{2s})
\end{equation}
The Mayer transfer operator for $PGL(2,\mathbb Z)$, is a transfer operator corresponding to the dynamical system $\mathcal D_2$,
\begin{equation}\label{transfer-Gx}
\mathcal L_sf(x)=\sum_{n=1}^\infty (\dfrac{1}{x+n})^{2s}f(\dfrac{1}{x+n})
\end{equation} 
which is sometimes called the transfer operator for the billiard flow or for the Gauss map because the dynamics of $\mathcal D_2$ is defined by the Gauss map $T_G$. The analytic properties of this operator easily extend to the original Mayer operator. Thus from now on we consider this simplified version of the Mayer operator.

Up to now we have seen how the ergodic aspects of the geodesic flow fix the form of the Mayer operator with no information about the function space on which $\mathcal L_s$ acts . In the next step we should decide about this space in such a way that the operator $\mathcal L_s$ becomes a nice operator from an analytic point of view with well defined trace and determinant. To do this we need first two lemmas. Let the real $x$ in (\ref{transfer-Gx}) be extended to $z$ in the complex domain \begin{equation}
D_r=\left\lbrace z\in\mathbb C \ \vert \ \vert z-1\vert<r\right\rbrace,
\end{equation}
\begin{lem1}
For $n\in \mathbb N$, the map $\psi_n:z\rightarrow\frac{1}{z+n}$  maps holomorphically the disc $D_r$ strictly inside itself if the radius $r$ of the disc lies in the interval $(1,\frac{1+\sqrt{5}}{2})$. For $r=1$ the map $\psi_n$ touches the boundary in the limit $n\rightarrow\infty$.
\end{lem1}
\begin{proof}
First we note that 
\begin{equation}
\lim_{n\rightarrow\infty}\psi_n(z)=0, \ \ \ z\in D_r,
\end{equation}
consequently the smallest lower bound of the radius is $r=1$. But the upper bound is at most $r<2$ because otherwise $\psi_1$ is not contracting at $z=-1$. To determine the largest upper bound of $r$  we note that the maps $\psi_n$'s are conformal, mapping circles into circles. We also note that the $\psi_n$'s map the disc $D_r$ to discs with the centre on the real axis. Thus to investigate the contracting properties of $\psi_n$ it is enough to investigate the transformations of the end points $x_-=1-r$ or $x_+=1+r$ of the diameter of $D_r$ lying on the real axis. Moreover because of the fact that
\begin{equation}
\psi_{n+1}(x)<\psi_n(x), \ \ \ \ n\in\mathbb N,\ \ \ x>-1,
\end{equation}
to find the upper bound of $r$, it is enough to find the upper bound of $r$ for $\psi_1$. But
\begin{equation}
\psi_1(x_+)<\psi_1(x_-),\ \ \ \ 1<r<2
\end{equation}
hence the largest upper bound of the radius is determined by the following inequality
\begin{equation}
\psi_1(x_-)=\dfrac{1}{2-r}<1+r.
\end{equation}
By simple calculation we get $r<\frac{1+\sqrt{5}}{2}$ which completes the proof.
\end{proof}
The following lemma suggests a possible good candidate for the desired space. 
\begin{lem2}
For $r\in(1,\frac{1+\sqrt{5}}{2})$, the Banach space
\begin{equation}
B(D_r)=\left\lbrace f:D_r\rightarrow\mathbb C \ \vert \ f \ \text{holomorphic on \ $D_r$\ and continuous on} \ \overline{D}_r\right\rbrace 
\end{equation}
with the sup norm $\vert \vert \ \vert\vert$ is invariant under the action of $\mathcal L_s$ for $Re(s)>\dfrac{1}{2}$, that is $\mathcal L_sB(D_r)\subset B(D_r)$. Note that the continuity on the boundary makes the space $B(D_r)$ a subspace of holomorphic bounded functions.
\end{lem2}
\begin{proof}
The transfer operator for $PGL(2,\mathbb Z)$ is given by 
\begin{equation}\label{transfer-Gz}
\mathcal L_sf(z)=\sum_{n=1}^\infty (\dfrac{1}{z+n})^{2s}f(\dfrac{1}{z+n})
\end{equation}
Let $f$ be in $B(D_r)$. The argument of $f$, that is the function $\psi_n(z)=\dfrac{1}{z+n}$, maps the disc $D_r$ inside itself and is holomorphic in this domain for all $n\in \mathbb N$. Thus the term $S_n:=(\dfrac{1}{z+n})^{2s}f(\dfrac{1}{z+n})$ also belongs to $B(D_r)$. We are going to prove that $\sum_{n=1}^\infty S_n\in B(D_r)$. The required result comes from the general Weierstrass M-test, for which we need $\vert \vert S_n \vert\vert\leq M_n<\infty$ for all $n\geq 1$ and $\sum_{n=1}^\infty M_n<\infty$ where $M_n$ is a positive number. These requirements are fulfilled for $\sigma=Re(s)>\frac{1}{2}$ if we take $M_n=(n-r+1)^{-2\sigma}\vert\vert f\vert\vert$ where $\vert\vert f\vert\vert$ is bounded because $f\in B(D_r)$. This completes the proof. 
\end{proof}
\begin{cor1}
Let $r$ lies in the interval $[1,\frac{1+\sqrt{5}}{2})$. Then the operator $\overset{\sim}{\mathcal L}_s$ for the group $PSL(2,\mathbb Z)$ is defined in the Banach space $B(D_r)\oplus B(D_r)$ for $Re(s)>\frac{1}{2}$ and leaves this space invariant.
\end{cor1}
In subsection 2.2 we will see that this choice of the function space makes the Mayer operator a nuclear operator which confirms that we are on the right track.
\subsection{Matrix representation of the transfer operator, its eigenvectors and eigenvalues}
We will see in the next subsection that the transfer operator is  compact and the spectrum of this operator in the space $B(D_r)$ is a discrete set of eigenvalues of finite multiplicity. We are going to derive now a matrix representation for the transfer operator.

Let $f\in B(D_r)$ be an eigenfunction of $\mathcal L_s$, 
\begin{equation}\label{eigen-eq}
\mathcal L_s f(z)=\lambda f(z),
\end{equation}
the holomorphy of $f$ on $D_r$ allows the following expansion,
\begin{equation}\label{ffff}
f(z)=\sum_{k=0}^\infty c_k (z-1)^k, \ \ \ \ z\in D_r.
\end{equation}
Inserting this expansion into (\ref{eigen-eq}) we get,
\begin{equation}\label{pp}
\sum_{n=1}^\infty(\dfrac{1}{z+n})^{2s}\sum_{k=0}^\infty c_k(\dfrac{1}{z+n}-1)^k=\lambda \sum_{k=0}^\infty c_k (z-1)^k
\end{equation}
but
\begin{equation}
(\dfrac{1}{z+n}-1)^k=\sum_{j=0}^k\binom{k}{j}(-1)^{k-j}(\dfrac{1}{z+n})^j
\end{equation}
therefore (\ref{pp}) can be written as
\begin{equation}\label{pp12}
\sum_{k=0}^\infty c_k\sum_{j=0}^k\binom{k}{j}(-1)^{k-j}\sum_{n=1}^\infty(\dfrac{1}{z+n})^{2s+j}=\lambda \sum_{k=0}^\infty c_k (z-1)^k
\end{equation}
On the other hand we have the following Taylor expansion at $z=1$,
\begin{equation}
(\dfrac{1}{z+n})^{2s+j}=\sum_{m=0}^\infty \dfrac{(-1)^m}{m!}\dfrac{\Gamma(2s+j+m)}{\Gamma(2s+j)}(\dfrac{1}{1+n})^{2s+j+m}(z-1)^m
\end{equation}
thus we get,
\begin{equation}\label{ppp}
\sum_{n=1}^\infty(\dfrac{1}{z+n})^{2s+j}=\sum_{m=0}^\infty \dfrac{(-1)^m}{m!}\dfrac{\Gamma(2s+j+m)}{\Gamma(2s+j)}(\zeta(2s+j+m)-1)(z-1)^m
\end{equation}
where we used the following identity
\begin{equation}
\sum_{n=1}^\infty(\dfrac{1}{1+n})^{\beta}=\zeta(\beta)-1.
\end{equation}
Finally by inserting (\ref{ppp}) into (\ref{pp12}) we get,
\begin{equation}\label{qqqq}
\sum_{k=0}^\infty\sum_{m=0}^\infty a_{mk}(s)c_kz^m=\lambda \sum_{k=0}^\infty c_k z^k
\end{equation}
where
\begin{equation}\label{array}
a_{mk}(s)=\sum_{j=0}^k\binom{k}{j}\dfrac{(-1)^{m+k-j}}{m!}\dfrac{\Gamma(2s+j+m)}{\Gamma(2s+j)}(\zeta(2s+j+m)-1)
\end{equation}
which finally leads to the following eigenvalue equation for the transfer operator in the matrix representation,
\begin{equation}\label{eigen-eq-mat}
\sum_{k=0}^\infty a_{mk}(s)c_k=\lambda c_m
\end{equation}
Based on these calculations we can formulate the following lemma
\begin{matrix-rep}
In the natural basis, $\left\lbrace z^k\right\rbrace_{k=0}^\infty$, according to (\ref{ffff}), the eigenfunctions $f\in B(D_r)$ have the representation given by the sequence $\left\lbrace c_k\right\rbrace_{k=0}^\infty$ which satisfies (\ref{eigen-eq-mat}). Moreover the transfer operator in this basis is an infinite dimensional matrix whose matrix elements are given by (\ref{array}). 
\end{matrix-rep}  
\begin{matrix-rep1}
Mayer \cite{Mayer90} derived a simpler matrix representation of transfer operator given by,
\begin{equation}
a_{mk}(s)=\dfrac{(-1)^{m}}{m!}\dfrac{\Gamma(2s+k+m)}{\Gamma(2s+k)}\zeta(2s+k+m).
\end{equation}
In this representation the basis of the space $B(D_r)$ is chosen as,
\begin{equation}
\left\lbrace \zeta(2s+k,z+1)\right\rbrace_{k=0}^\infty
\end{equation}
where $\zeta(s,w)$ denotes the Hurwitz zeta function.
\end{matrix-rep1}
\subsection{Nuclear Spaces, Nuclear Operators and Grothendieck theory}
Let $B$ be an arbitrary Banach space and denote its  dual by $B^*$ which is the space of bounded functionals on $B$. A linear operator $\mathcal L$ on $B$ is called nuclear of order $q$ if it has the representation (see \cite{mayer book})
\begin{equation}\label{nuclear-op}
\mathcal L=\sum_n \lambda_nf^*_n(.)f_n
\end{equation}
where $\left\lbrace f_n\right\rbrace $ and $\left\lbrace f^*_n\right\rbrace$ are families in $B$ and $B^*$ respectively with $\vert\vert f_n\vert\vert\leq 1$ and $\vert\vert f^*_n\vert\vert\leq 1$. Furtheremore $\left\lbrace \lambda_n\right\rbrace $ is a sequence of complex numbers and $q=\inf\left\lbrace \epsilon\leq1 \ \vert \ \sum_n\vert \lambda_n\vert^\epsilon < \infty \right\rbrace$.

For certain classes of nuclear operators a trace functional is available. According to a remarkable result of Grothendieck, nuclear operators $\mathcal L$ of order $q\leq\frac{2}{3}$ have a trace $tr \mathcal =\sum \rho_i$ as the sum of eigenvalues $\rho_i$ counted with multiplicities. Moreover the Fredholm determinant of $\mathcal L$ is defined as
\begin{equation}
det(1-z\mathcal L)=exp \ tr \ log(1-z\mathcal L)=\prod_i(1-\rho_iz)
\end{equation}
or equivalently
\begin{equation}\label{def-Gro-det}
det(1-z\mathcal L)=exp(-\sum_{n=1}^\infty\frac{z^n}{n}tr\mathcal L^n).
\end{equation}
Consequentely if $\mathcal L=\mathcal L(s)$ is a holomorphic function of a parameter $s$ in a domain then the corresponding determinant is also a holomorphic function of $s$ in the given domain (see \cite{Gro,mayer book}). 
\begin{lem3}
The Mayer transfer operator $\mathcal L_s$ for the group $PGL(2,\mathbb Z)$, given by 
\begin{equation}\label{formula for this lemma}
\mathcal L_sf(z)=\sum_{n=1}^\infty (\dfrac{1}{z+n})^{2s}f(\dfrac{1}{z+n}),
\end{equation}
acting on the Banach space $B(D_r)$, defined as in lemma.2, in the domain $\sigma=Re(s)>\dfrac{1}{2}$, is a nuclear operator of order $q=\dfrac{1}{2\sigma}$.
\end{lem3}
\begin{proof}
We are going to find a representation of Mayer transfer operator in the form of (\ref{nuclear-op}). To avoid overloading we give the proof for $r=\frac{3}{2}$. First we insert the Taylor expansion of $f$ at $z=1$ in (\ref{formula for this lemma}),
\begin{equation}
\mathcal L_sf(z)=\sum_{n=1}^\infty\sum_{m=0}^\infty\dfrac{f^{(m)}(1)}{m!}(\dfrac{1}{z+n}-1)^m(\dfrac{1}{z+n})^{2s}.
\end{equation}
We introduce the following family of functions,
\begin{equation}
f_{n,m}(z):=(\dfrac{1}{a_n})^{2\sigma}(\dfrac{1}{z+n}-1)^m(\dfrac{1}{z+n})^{2s}
\end{equation}
where 
\begin{equation}
a_n:=\sup_{z\in D_r}\dfrac{1}{z+n}=\dfrac{1}{n-\frac{1}{2}}.
\end{equation}
We note that 
\begin{equation}
\sup_{z\in D_r}\vert\dfrac{1}{z+n}-1\vert\leq1
\end{equation}
thus we have
\begin{equation}
\Vert f_{n,m}\Vert\leq1.
\end{equation}
Next we define
\begin{equation}
f^*_{n,m}(f):=r^m\dfrac{f^{(m)}(1)}{m!}.
\end{equation}
Obviously according to Cauchy estimates we have
\begin{equation}
\Vert f^*_{n,m}\Vert\leq1.
\end{equation}
The desired representation is given by 
\begin{equation}
\mathcal L_sf(z)=\sum_{n=1}^\infty\sum_{m=0}^\infty \lambda_{n,m}f^*_{n,m}(f)f_{n,m}(z)
\end{equation}
with
\begin{equation}
\lambda_{n,m}:=r^{-m}(a_n)^{2\sigma}.
\end{equation}
We note that
\begin{equation}
\sum_{n=1}^\infty\sum_{m=0}^\infty\vert\lambda_{n,m}\vert^\epsilon=\sum_{m=0}^\infty r^{-\epsilon m}\sum_{n=1}^\infty(\dfrac{1}{n-\frac{1}{2}})^{2\epsilon\sigma}.
\end{equation}
The first sum is a geometrical series, absolutely convergent for any $\epsilon>0$ and the second sum for any $\epsilon>\dfrac{1}{2\sigma}$ is absolutely convergent to $(2^{2\sigma\epsilon}-1)\zeta(2\sigma\epsilon)$.
Therefore we have
\begin{equation}
\sum_{n=1}^\infty\sum_{m=0}^\infty\vert\lambda_{n,m}\vert^\epsilon=\dfrac{1}{1-(r^{-1})^\epsilon}(2^{2\sigma\epsilon}-1)\zeta(2\sigma\epsilon), \ \ \ \ \epsilon>\dfrac{1}{2\sigma}.
\end{equation}
Our later discussion shows that for $\sigma>\dfrac{1}{2}$,
\begin{equation}
\inf \left\lbrace \epsilon \ \vert \ \sum_{n=1}^\infty\sum_{m=0}^\infty\vert\lambda_{n,m}\vert^\epsilon<\infty\right\rbrace=\dfrac{1}{2\sigma}
\end{equation}
which completes the proof.
\end{proof}
Mayer proved a stronger assertion about the nuclearity of the transfer operator in a more elegant way (see \cite{Mayer76}). In fact he proved that the transfer operator is nuclear of order zero. This proof is based on some properties of so called nuclear spaces. 
It was Alexander Grothendieck who first introduced the class of nuclear spaces (see \cite{Gro}). Roughly speaking nuclear spaces are the maximal class of linear topological spaces with nice properties from analytic point of view. For example they admit a generalized kernel theorem of L. Schwartz. More presicely (see \cite{Sch}, page 100),
\begin{defdd}
A locally convex topological vector space $E$ is a nuclear space if and only if every continuous linear map of $E$ into any Banach space is nuclear.
\end{defdd}
The space $H(D)$ of holomorphic functions on the open disc $D$ with the open compact topology is an example of a nuclear space. The space $H(D)$ fulfills stronger assertions than that of deffinition.1. Indeed, every bounded linear map of the nuclear space $H(D)$ to any Banach space is not only nuclear but nuclear of order zero \cite{Gro}. For more details about the nuclear spaces we refer to \cite{Gro,Gel4,Sch}. 

Now, to prove the nuclearity of the transfer operator, we extend it first to the nuclear space $H(D_r)\supset B(D_r)$ with $r\in (1,\frac{1+\sqrt{5}}{2})$, that is 
\begin{eqnarray}
&\widehat{\mathcal L}_s:H(D_r)\rightarrow B(D_r)&\nonumber\\&\widehat{\mathcal L}_sf(z)=\sum_{n=1}^\infty (\dfrac{1}{z+n})^{2s}f(\dfrac{1}{z+n}),&
\end{eqnarray}
\begin{lem4}
The operator $\widehat{\mathcal L}_s:H(D_r)\rightarrow B(D_r)$ for $Re(s)>\frac{1}{2}$ is nuclear of order zero where $r\in (1,\frac{1+\sqrt{5}}{2})$.
\end{lem4}
\begin{proof}
As mentioned before, every bounded linear map of $H(D_r)$ to a Banach space is nuclear of order zero. Thus it is enough to prove the boundedness of $\widehat{\mathcal L}_s$. To this end we should show that there exists a neighbourhood of zero $V(0)\subset H(D_r)$ which is mapped to a bounded subset of $B(D_r)$. To begin with we define the following sequence of open discs,
\begin{equation}
\mathbb K_n=\left\lbrace w=\psi_n(z) \ \vert \ z\in D_r\right\rbrace 
\end{equation}
whose radius $r_n$ and centre $c_n$ is given by
\begin{equation}
r_n=\dfrac{r}{(n+1)^2-r^2}
\end{equation}
and
\begin{equation}
c_n=(\dfrac{n+1}{(n+1)^2-r^2},0).
\end{equation}
We choose the following neighbourhood of zero $V(0)\subset H(D_r)$,
\begin{equation}
V(0)=\left\lbrace f\in H(D_r)\ \vert \ \sup_{z\in \overline{\mathbb K}_{1}}\vert f(z)\vert<M\right\rbrace 
\end{equation}
where $M>0$ is a constant.
Also the following holds
\begin{equation}\label{elegant way}
\mathbb K_{n}\subset\overline{\mathbb K}_{1}, \ \ \ n\in \mathbb N.
\end{equation}
For any $f\in H(D_r)$ we have 
\begin{equation}\label{qo}
\sup_{z\in D_r}\vert \widehat{\mathcal L}_sf(z)\vert\leq\sum_{n=1}^\infty( \dfrac{1}{1-r+n})^{2\sigma}\sup_{z\in D_r}f(\psi_n(z)).
\end{equation}
But because of (\ref{elegant way}),
\begin{equation}
\sup_{z\in D_r}f(\psi_n(z))\leq\sup_{z\in \mathbb K_{1}}f(z)=M.
\end{equation}
Inserting this into (\ref{qo}) completes the proof. 
\end{proof}
On the other hand we have \begin{displaymath}
\mathcal L_s=\widehat{\mathcal L}_s\circ\imath
\end{displaymath}
where $\imath$ is the bounded injection given by
\begin{displaymath}
\imath:B(D_r)\rightarrow H(D_r), \ \ \ \ \imath(f)=f.
\end{displaymath}
Then nuclearity of $\widehat{\mathcal L}_s$ leads to nuclearity of $\mathcal L_s$ because of the fact that the product of a bounded operator with a nuclear operator is also nuclear with the same order. Thus we have the following corrolary,
\begin{cor2}
The Mayer transfer operator for the groups $PGL(2,\mathbb Z)$ and $PSL(2,\mathbb Z)$ in the domain $Re(s)>\frac{1}{2}$ is nuclear of order zero. 
\end{cor2}
Note that in this corrolary we concluded the nuclearity of the transfer operator for $PSL(2,\mathbb Z)$ from the nuclearity of the transfer operator for $PGL(2,\mathbb Z)$ and (\ref{Mayer vect}).

One of the consequences of the nuclearity of an operator is its compactness (see \cite{Sch}, page 99). Thus we attain the following corollary,
\begin{cor3}
The Mayer transfer operators for both of the groups $PSL(2,\mathbb Z)$ and $PGL(2,\mathbb Z)$ in the domain $Re(s)>\frac{1}{2}$ are compact.
\end{cor3}
\section{Integral representation of Mayer transfer operator}
In this section we discuss a new model for the Mayer transfer operator in a Hilbert space. This is important for the investigation of the transfer operator because in this model the Grothendieck theory of nuclear operator in Banach spaces is reduced to the simpler theory of linear operators in Hilbert spaces. 

We follow Mayer \cite{mayer book} to derive an integral representation for the transfer operator in a Hilbert space for $Re(s)>\frac{1}{2}$. Let $\mathcal J_{\nu}(u)$ denotes the Bessel function (see \cite{Gradshteyn}),
\begin{equation}\label{GGGG}
\mathcal J_\nu(u)=\sum_{k=0}^\infty(\frac{u}{2})^{2k+\nu}\dfrac{(-1)^k}{k!\Gamma(k+\nu+1)}
\end{equation}
Then it is not difficult to check the validity of the following inequality 
\begin{equation}
\int_0^\infty\int_0^\infty \vert\mathcal J_{2s-1}(2\sqrt{tt'})\vert^2\frac{dt}{e^t-1}\frac{dt'}{e^{t'}-1}<\infty \ \ \ \ Re(s)>\frac{1}{2}
\end{equation}
Thus in the domain $Re(s)>\frac{1}{2}$, the Bessel function $\mathcal J_{2s-1}(2\sqrt{tt'})$ can be considered as a Hilbert-Schmidt kernel with respect to the measure
\begin{equation}\label{measure}
dm(t)=\frac{dt}{e^t-1}
\end{equation}
Therefore we can define a Hilbert-Schmidt integral operator $\mathcal K_s $ given by
\begin{equation}
\mathcal K_s\varphi(t)=\int_0^\infty\mathcal J_{2s-1}(2\sqrt{tt'})\varphi(t') dm(t') \ \ \ \ \varphi\in L_2(\mathbb R^+,dm)
\end{equation}
where $L_2(\mathbb R^+,dm)$ denotes the Hilbert space of square integrable functions on the positive real axis with respect to the measure $dm$ given in (\ref{measure}). The operator $\mathcal K_s$ is obviously bounded in this space for $Re(s)>\frac{1}{2}$. 

We are going to explain now how this operator is related to the Mayer transfer operator. First we consider the following integral transform,
\begin{equation}\label{phi-transform}
(T_s\varphi)(z)=\int_0^\infty e^{-zt}t^{s-\frac{1}{2}}\varphi(t)dm(t) \ \ \ \ \varphi\in L_2(\mathbb R^+,dm) \ \ \ \ z\in\mathbb C/[-1,\infty)
\end{equation}
This transform can be considered as a composition of a multiplication by the exponential function and a Mellin transform, both of them are formally invertible thus the transform $T_s$ is formally invertible. For a complex $s$ with $Re(s)>\frac{1}{2}$, consider the space $\mathcal H_1$ of holomorphic functions in the domain $\mathbb C/[-1,\infty)$ which is the image of $L_2(\mathbb R^+,dm)$ under $T_s$,
\begin{equation}
\mathcal H_1=\left\lbrace f(z)=(T_s\varphi)(z)\ \vert \ \varphi\in L_2(\mathbb R^+,dm),\ z\in\mathbb C/[-1,\infty)\right\rbrace 
\end{equation}
The operator 
\begin{equation}
\mathcal L'_s:\mathcal H_1\rightarrow \mathcal H_1
\end{equation}
given by
\begin{equation}
\mathcal L'_s=T_s\mathcal K_s T_s^{-1}
\end{equation}
is isomorphic to $\mathcal K_s$, with the same spectrum.
Shortly we see that $\mathcal L'_s$ has the same form as Mayer's operator.
\begin{lem5}
For $Re(s)>\frac{1}{2}$ on the space $\mathcal H_1$, the following identity holds,
\begin{equation}
\mathcal L'_s f(z)=\sum_{n=1}^\infty (\dfrac{1}{z+n})^{2s}f(\dfrac{1}{z+n}).
\end{equation}
\end{lem5}
\begin{proof}
Let $f(z)=(T_s\varphi)(z)$. Then
\begin{equation}
\mathcal L'_s f(z)=T_s(z)\mathcal K_s\varphi=\int_0^\infty dm(t) e^{-zt}t^{s-\frac{1}{2}}(\mathcal K_s\varphi)(t)
\end{equation}
or 
\begin{equation}\label{step}
\mathcal L'_s f(z)=\int_0^\infty dt\dfrac{t^{s-\frac{1}{2}}}{e^t-1}e^{-zt}\int_0^\infty \mathcal J_{2s-1}(2\sqrt{tt'})\varphi(t')dm(t') 
\end{equation}
From (\ref{GGGG}) we have
\begin{equation}\label{B-expand}
\mathcal J_{2s-1}(2\sqrt{tt'})=(tt')^{s-\frac{1}{2}}\sum_{k=0}^\infty\dfrac{(-tt')^k}{k!\Gamma(k+2s)}
\end{equation}
Inserting (\ref{B-expand}) into (\ref{step}) and rearranging the terms we get,
\begin{equation}
\mathcal L'_s f(z)=\int_0^\infty t'^{s-\frac{1}{2}}\varphi(t')dm(t') \int_0^\infty dt\dfrac{t^{2s-1}}{e^t-1}e^{-zt}\sum_{k=0}^\infty\dfrac{(-tt')^k}{k!\Gamma(k+2s)}
\end{equation}
or
\begin{equation}\label{step2}
\mathcal L'_s f(z)=\int_0^\infty t'^{s-\frac{1}{2}}\varphi(t')dm(t') \sum_{k=0}^\infty\dfrac{(-t')^k}{k!\Gamma(k+2s)}\int_0^\infty dt\dfrac{t^{2s+k-1}e^{-zt}}{e^t-1}
\end{equation}
But the Hurwitz zeta function has the integral presentation \cite{Gradshteyn},
\begin{equation}
\zeta(w;q)=\frac{1}{\Gamma(w)}\int_0^\infty\dfrac{t^{w-1}e^{-qt}}{1-e^{-t}}dt \ \ \ \ Re(w)>1
\end{equation}
That is the integral over $t$ in (\ref{step2}) can be replaced by the Hurwitz zeta function,
\begin{equation}
\mathcal L'_s f(z)=\int_0^\infty t'^{s-\frac{1}{2}}\varphi(t')dm(t') \sum_{k=0}^\infty\dfrac{(-t')^k}{k!}\zeta(k+2s;z+1)
\end{equation} 
On the other hand for $Re(w)>1$ the Hurwitz zeta function is defined as
\begin{equation}
\zeta(w;q)=\sum_{n=0}^\infty(\frac{1}{q+n})^w
\end{equation}
which leads to the following formula
\begin{equation}
\mathcal L'_s f(z)=\int_0^\infty t'^{s-\frac{1}{2}}\varphi(t')dm(t') \sum_{k=0}^\infty\dfrac{(-t')^k}{k!}\sum_{n=1}^\infty(\frac{1}{z+n})^{2s+k}
\end{equation} 
or
\begin{equation}
\mathcal L'_s f(z)=\sum_{n=1}^\infty(\frac{1}{z+n})^{2s}\int_0^\infty t'^{s-\frac{1}{2}}\varphi(t')dm(t') \sum_{k=0}^\infty\dfrac{(-t')^k}{k!}(\frac{1}{z+n})^k
\end{equation} 
The summation over $k$ is nothing than the Taylor expansion of the function $e^{-t'(\frac{1}{z+n})}$, thus we get
\begin{equation}
\mathcal L'_s f(z)=\sum_{n=1}^\infty(\frac{1}{z+n})^{2s}\int_0^\infty t'^{s-\frac{1}{2}}e^{-t'(\frac{1}{z+n})}\varphi(t')dm(t') 
\end{equation}
or according to (\ref{phi-transform})
\begin{equation}
\mathcal L'_s f(z)=\sum_{n=1}^\infty(\frac{1}{z+n})^{2s}(T_s\varphi)(\frac{1}{z+n})
\end{equation}
That means
\begin{equation}
\mathcal L'_s f(z)=\sum_{n=1}^\infty(\frac{1}{z+n})^{2s}f(\frac{1}{z+n})
\end{equation}
\end{proof}
As we saw in this lemma the Mayer transfer operator $\mathcal L_s$ and the operator $\mathcal L'_s$ has the same form. From this fact and regarding the spaces on which these operators act, we can see that every eigenfunction of $\mathcal L'_s$ is an eigenfunction of $\mathcal L_s$. In \cite{Mayer90}, Mayer proved the converse, thus we achieve the following lemma
\begin{LL'}
For $Re(s)>\frac{1}{2}$, the operators $\mathcal L'_s$ and $\mathcal L_s$ have the same spectrum.
\end{LL'}
\begin{cor4}
In the domain $Re(s)>\frac{1}{2}$, the integral operator $\mathcal K_s$ and the Mayer transfer operator for $PGL(2,\mathbb Z)$, $\mathcal L_s$ have the same spectrum counted with multiplicities.
\end{cor4}
\section{Calculation of the trace}
As explained in section $2$, the transfer operator is of the trace class. In this part we are going to calculate the trace of the transfer operator and its powers. We illustrate two different approaches for the calculation of trace, the first one based on the contracting property of the map 
\begin{equation}\label{map}
\psi_n(z)=\dfrac{1}{z+n}
\end{equation}
which allows to apply the method of geometric trace and the second one uses the integral representation of the transfer operator.

By using (\ref{map}), the Mayer transfer operator for $PGL(2,\mathbb Z)$ can be written in the form
\begin{equation}
\mathcal L_s f(z)=\sum_{n=1}^\infty\psi_n(z)^{2s}f(\psi_n(z))
\end{equation}
To calculate the trace of $\mathcal L_s$, first we need to calculate the trace of the terms,
\begin{equation}
\mathcal L_{s,n}f(z)=\psi_n(z)^{2s}f(\psi_n(z)) \ \ \ \ n\in\mathbb N
\end{equation}
With the same arguments as in section 2.1, one can see that the operator $\mathcal L_{s,n}$ is nuclear of order zero for all $n\in \mathbb N$. For $r\in [1,\frac{1+\sqrt{5}}{2})$ the map $\psi_n(z)$ on $D_r$ has a unique fixed point $z_n^*$ given by
\begin{equation}\label{fixed point}
z_n^*=-\frac{n}{2}+\frac{\sqrt{n^2+4}}{2}
\end{equation}
which obtains by solving an equation
\begin{equation}\label{fix-p-eq}
\dfrac{1}{z+n}=z.
\end{equation}

The existence of a unique solution for (\ref{fix-p-eq}) in $D_r$ is crucial for the calculation of the trace of $\mathcal L_{s,n}$. Before proceeding furthere we quote a Lemma from \cite{mayer book} concerning the eigenvalues of a general composition operator.
\begin{lem0}
Let $D\subset\mathbb C$ be an arbitrary domain, $\psi$ be a map on $D$ with a unique fixed point $z^*\in D$ and $\varphi$ be an arbitrary function in the Banach space $B(D)$. Then the spectrum of the composition operator $\mathcal Cf=\varphi \psi\circ f$ on $B(D)$ consists of simple eigenvalues $\lambda_n=\varphi(z^*)(\psi'(z^*))^n, \ n=0, 1, ...$ which converge to zero as $n\rightarrow\infty$.  
\end{lem0}
\begin{rem2}
A contracting map $\psi$ on a domain $D$, is said to be a map of the domain $D$ strictly inside itself if,
\begin{equation}
\underset{z\in D,z'\in  \mathbb C\backslash D}{inf}\Vert \psi(z)-z'\Vert\geq\delta>0
\end{equation}
Such maps have allways a unique fixed point in $D$.
\end{rem2}
According to this lemma the eigenvalues of $\mathcal L_{s,n}$ are given by
\begin{equation}
\lambda_m(n)=(\psi_n(z_n^*))^{2s}(\psi'_n(z_n^*))^{m}=(-1)^m(z_n^*)^{2s}(z_n^*)^{2m},\ \ \ \ m\in\mathbb N\cup\left\lbrace 0 \right\rbrace 
\end{equation}
which are all simple. Therefore the trace of $\mathcal L_{s,n}$ is simply the sum of them,
\begin{equation}
tr\mathcal L_{s,n}=\sum_{m=0}^\infty \lambda_m(n)= \dfrac{(z_n^*)^{2s}}{1+(z_n^*)^2}
\end{equation}
Then we attain the trace of $\mathcal L_s$ for $Re(s)>\frac{1}{2}$ by summing the traces of all $\mathcal L_{s,n}$'s,
\begin{equation}\label{tr1356}
tr\mathcal L_s=\sum_{n=1}^\infty \dfrac{(z_n^*)^{2s}}{1+(z_n^*)^2}
\end{equation}
where $Re(s)>\frac{1}{2}$ ensures the absolute convergence.

Next we calculate the trace of the powers of the transfer operator. First we note that 
\begin{equation}\label{sum-comp}
\mathcal L_s^n=\sum_{i_1\geq1}\sum_{i_2\geq1}...\sum_{i_n\geq1}\mathcal L_{s,i_1}\mathcal L_{s,i_2}...\mathcal L_{s,i_n}
\end{equation}
where as before 
\begin{equation}
\mathcal L_{s,i_k}f(z)=\psi_{i_k}(z)^{2s}f(\psi_{i_k}(z))
\end{equation}
and the composition operator $\mathcal L_{s,i_1}\mathcal L_{s,i_2}...\mathcal L_{s,i_n}$ has the form,
\begin{eqnarray}\label{fff}
&\mathcal L_{s,i_1}\mathcal L_{s,i_2}...\mathcal L_{s,i_n}f(z)=&\nonumber\\& \left\lbrace \psi_{i_1}(z)[\psi_{i_2}\psi_{i_1}(z)].\ .\ .[\psi_{i_n}\psi_{i_{n-1}}...\psi_{i_1}(z)]\right\rbrace ^{2s}f(\psi_{i_n}...\psi_{i_1}(z))& 
\end{eqnarray}

For the sake of the convenience, without fear of confusion we introduce the following notations, 
\begin{equation}\label{def-comp}
\mathcal L_{s,\underline{n}}:=\mathcal L_{s,i_1}\mathcal L_{s,i_2}...\mathcal L_{s,i_n}\ \ \ \psi_{\underline{n}}:=\psi_{i_n}...\psi_{i_1}(z)
\end{equation}
by which formula (\ref{fff}) can be written as the following
\begin{equation}\label{com}
\mathcal L_{s,\underline{n}}f(z)=\left\lbrace \prod_{k=1}^n\psi_{\underline{k}}(z)\right\rbrace ^{2s}f(\psi_{\underline{n}}(z))
\end{equation}
On the other hand the composition map $\psi_{\underline{n}}$ has a unique fixed point, $z_{\underline{n}}$ on $D_r$ given by a periodic continuous fraction,
\begin{equation}\label{fp}
z_{\underline{n}}=[0,\overline{i_n,i_{n-1},...,i_1}]
\end{equation}
The uniqueness of the fixed point $z_{\underline{n}}$ enables us to apply lemma.8 for the composition operator $\mathcal L_{s,\underline{n}}$ in (\ref{com}). Therefore we achieve the eigenvalues of $\mathcal L_{s,\underline{n}}$ given by
\begin{equation}\label{e-v}
\lambda_m(z_{\underline{n}})=\left\lbrace \prod_{k=1}^n\psi_{\underline{k}}(z_{\underline{n}})\right\rbrace ^{2s}(\frac{d\psi_{\underline{n}}}{dz}\vert_{z=z_{\underline{n}}})^{m},\ \ \ m\in\mathbb N
\end{equation}
By using the chain rule and noting that 
\begin{equation}
\frac{d\psi_{i_k}}{dz}=(-1)(\psi_{i_k})^2, \ \ \ i_k\in\mathbb N
\end{equation}
the derivative of the composition function $\psi_{\underline{n}}(z)$ can be written as
\begin{equation}\label{deriv}
\frac{d}{dz}\psi_{\underline{n}}(z)\vert_{z=z_{\underline{n}}}=(-1)^n(\prod_{k=1}^n\psi_{i_k}\psi_{\underline{k-1}}(z_{\underline{n}}))^{2}=(-1)^n(\prod_{k=1}^n\psi_{\underline{k}}(z_{\underline{n}}))^{2}
\end{equation}
where $\psi_{\underline{0}}:=id$ and the last identity simply comes from definition of $\psi_{\underline{k}}$ in (\ref{def-comp}). By inserting (\ref{deriv}) into (\ref{e-v}), the eigenvalues $\lambda_m(z_{\underline{n}})$ can be written as the following
\begin{equation}\label{e-v2}
\lambda_m(z_{\underline{n}})=(-1)^{nm}\left\lbrace \prod_{k=1}^n\psi_{\underline{k}}(z_{\underline{n}})\right\rbrace ^{2s+2m},\ \ \ m\in\mathbb N
\end{equation}
Moreover the following identity holds,
\begin{equation}
\psi_{\underline{k}}(z_{\underline{n}})=T^{k-1}z_{\underline{n}}
\end{equation}
where the Gauss map $T$ is given in (\ref{Gauss-map}). Thus the eigenvalues of $\mathcal L_{s,\underline{n}}$ in (\ref{e-v2}) can be written as
\begin{equation}
\lambda_m(z_{\underline{n}})=(-1)^{nm}\left\lbrace \prod_{k=1}^nT^{k-1}z_{\underline{n}}\right\rbrace ^{2s+2m},\ \ \ m\in\mathbb N
\end{equation}
or by transforming $k$ to $k+1$ in a simpler form,
\begin{equation}
\lambda_m(z_{\underline{n}})=(-1)^{nm}\left\lbrace \prod_{k=0}^{n-1}T^{k}z_{\underline{n}}\right\rbrace ^{2s+2m},\ \ \ m\in\mathbb N
\end{equation}
Because of the simplicity of the eigenvalues, the trace of $\mathcal L_{\underline{n}}$ is the sum of all $\lambda_m(z_{\underline{n}})$'s, 
\begin{equation}
tr \mathcal L_{s,\underline{n}}=\sum_{m=0}^\infty\lambda_m(z_{\underline{n}})=\dfrac{\left\lbrace \prod_{k=0}^{n-1}T^{k}z_{\underline{n}}\right\rbrace ^{2s}}{1-(-1)^{n}\left( \prod_{k=0}^{n-1}T^{k}z_{\underline{n}}\right)^2}
\end{equation}
or with a more explicit notations for $\mathcal L_{s,\underline{n}}$ and $z_{\underline{n}}$ given in (\ref{def-comp}) and (\ref{fp}) respectively,
\begin{equation}
tr \mathcal L_{s,i_1}\mathcal L_{s,i_2}...\mathcal L_{s,i_n}=\dfrac{\left\lbrace \prod_{k=0}^{n-1}T^{k}[0,\overline{i_n,i_{n-1},...,i_1}]\right\rbrace ^{2s}}{1-(-1)^{n}\left( \prod_{k=0}^{n-1}T^{k}[0,\overline{i_n,i_{n-1},...,i_1}]\right)^2}
\end{equation}
Finally, according to (\ref{sum-comp}) we attain the trace of $\mathcal L_s^n$ by summing the contribution of all $\mathcal L_{i_1}\mathcal L_{i_2}...\mathcal L_{i_n}$'s,
\begin{equation}\label{trLn}
tr \mathcal L_s^n=\sum_{i_1\geq1}\sum_{i_2\geq1}...\sum_{i_n\geq1}\dfrac{\left\lbrace \prod_{k=0}^{n-1}T^k[0,\overline{i_n,i_{n-1},...,i_1}]\right\rbrace ^{2s}}{1-(-1)^{n}\left( \prod_{k=0}^{n-1}T^{k}[0,\overline{i_n,i_{n-1},...,i_1}]\right)^2}
\end{equation}

We note that in formula (\ref{trLn}) the sum is over the set of all fixed points $[0,\overline{i_n,i_{n-1},...,i_1}]$ of the map $\psi_{\underline{n}}$ as mentioned in (\ref{fp}). But this set coincides with a subset of $Fix T^n$,
\begin{equation}\label{fix+}
Fix_+T^n=\left\lbrace x=[0,\overline{i_n,i_{n-1},...,i_1}] \ : i_1,...,i_{n-1},i_n\in \mathbb N\right\rbrace \subset FixT^n
\end{equation}
where
\begin{equation}
FixT^n=\left\lbrace x=[0,\overline{i_n,i_{n-1},...,i_1}] \ : i_1,...,i_{n-1},i_n\in \mathbb Z\right\rbrace 
\end{equation}
Thus (\ref{trLn}) can be written in a more compact form 
\begin{equation}\label{trLn2}
tr \mathcal L_s^n=\sum_{x\in Fix_+T^n}\dfrac{\left\lbrace \prod_{k=0}^{n-1}T^k(x)\right\rbrace ^{2s}}{1-(-1)^{n}\left( \prod_{k=0}^{n-1}T^{k}(x)\right)^2}
\end{equation}
\subsection{Calculation of trace via the integral representation}
In this subsection we calculate the trace of the integral operator for $Re(s)>\frac{1}{2}$
\begin{equation}
\mathcal K_s \varphi(t)=\int_0^\infty\mathcal J_{2s-1}(2\sqrt{tt'})\varphi(t') dm(t') \ \ \ \ \varphi\in L_2(\mathbb R^+,dm)
\end{equation}
According to standard results of theory of linear operators in Hilbert spaces the trace of $\mathcal K_s$ is given by the following integral
\begin{equation}
tr\mathcal K_s=\int_0^\infty\mathcal J_{2s-1}(2t) dm(t) 
\end{equation}
or by inserting the measure from (\ref{measure}),
\begin{equation}
tr\mathcal K_s=\int_0^\infty\dfrac{\mathcal J_{2s-1}(2t)}{e^t-1} dt 
\end{equation}
To calculate this integral we insert the following identity 
\begin{equation}
\dfrac{1}{e^t-1}=\sum_{n=1}^\infty e^{-nt}
\end{equation}
thus we have
\begin{equation}
tr\mathcal K_s=\sum_{n=1}^\infty\int_0^\infty e^{-nt}\mathcal J_{2s-1}(2t) dt 
\end{equation}
The integral above can be calculated,
\begin{equation}
\int_0^\infty e^{-nt}\mathcal J_{2s-1}(2t) dt=\dfrac{x_n^{2s}}{1+x_n^2} \ \ \ \ \  x_n=-\frac{n}{2}+\frac{\sqrt{n^2+4}}{2}
\end{equation}
Therefore 
\begin{equation}
tr\mathcal K_s=\sum_{n=1}^\infty\dfrac{x_n^{2s}}{1+x_n^2}
\end{equation}
which coincides with (\ref{tr1356}) that is,
\begin{equation}
tr\mathcal K_s=tr\mathcal L_s
\end{equation}
as we expected from corollary.4.
To calculate the trace of powers of $\mathcal K_s$ it is not difficult to see that
\begin{eqnarray}
&&tr\mathcal K_s^n =\nonumber\\&&\int_0^\infty dm(t_n)\ldots\int_0^\infty dm(t_1)\nonumber\\ &&\mathcal J_{2s-1}(2\sqrt{t_1t_2})\ldots \mathcal J_{2s-1}(2\sqrt{t_{n-1}t_{n}})\mathcal J_{2s-1}(2\sqrt{t_{n}t_{1}})
\end{eqnarray}
By calculating this integral we achieve the expected result 
\begin{equation}\label{di-eq}
tr\mathcal K_s^n=tr\mathcal L_s^n
\end{equation}
We don't know a simple direct proof of (\ref{di-eq})(see \cite{mayer book}, similar calculation has been done in \cite{Alexei2}). 
\section{Ruelle zeta function and transfer operator}
In this section we will denote different zeta functions by other letters not necessarily the $\zeta$. 

As it mentioned in the first section, the Ruelle zeta function for a given weighted dynamical system $(\Lambda,F,g)$ is defined by 
\begin{equation}\label{Ruelle zeta}
\zeta_R(z)=exp(\sum_{n=1}^\infty\dfrac{z^n}{n}\sum_{x\in FixF^n}\prod_{k=0}^{n-1}g(F^kx))
\end{equation}
The Ruelle zeta function for the dynamical system $\mathcal D_2$, defined in (\ref{dy-sy2}) at the point $z=1$, is reduced to a dynamical function $\xi(s)$, given by
\begin{equation}\label{zeta1}
\xi(s)=exp(\sum_{n=1}^\infty\dfrac{1}{n}\sum_{x\in Fix_+T^n}\prod_{k=0}^{n-1}(T^kx)^{2s})
\end{equation} 
In the following lemma we see the close connection of the Mayer transfer operator and the dynamical function $\xi(s)$.
\begin{lem7}
Let $\mathcal L_s$ be the Mayer transfer operator for $PGL(2,\mathbb Z)$ given in (\ref{formula for this lemma}) and $\xi(s)$ be the zeta function defined by (\ref{zeta1}). For $Re(s)>\frac{1}{2}$ the following identity holds
\begin{equation}
\dfrac{det(1+\mathcal L_{s+1})}{det(1-\mathcal L_s)}=\xi(s)
\end{equation}
where the determinant of the transfer operator is defined in the sense of Grothen-\\diek by (\ref{def-Gro-det}).
\end{lem7}
\begin{proof}
From lemma.4, we know that for $Re(s)>\frac{1}{2}$, the transfer operators $\mathcal L_s$ and $\mathcal L_{s+1}$ are both nuclear of order zero. Therefore according to (\ref{def-Gro-det}) we have,
\begin{equation}\label{dets1}
det(1+\mathcal L_{s+1})=exp(-\sum_{n=1}^\infty\frac{(-1)^n}{n}tr \mathcal L_{s+1}^n)
\end{equation} 
and
\begin{equation}\label{dets}
det(1-\mathcal L_s)=exp(-\sum_{n=1}^\infty\frac{1}{n}tr \mathcal L_s^n)
\end{equation}
Dividing (\ref{dets1}) by (\ref{dets}) we get,
\begin{equation}
\dfrac{det(1+\mathcal L_{s+1})}{det(1-\mathcal L_s)}=exp(\sum_{n=1}^\infty \frac{1}{n}\left\lbrace-(-1)^{n}tr \mathcal L_{s+1}^n+tr \mathcal L_{s}^n\right\rbrace ),\ \ \ \ Re(s)>\frac{1}{2}
\end{equation}
On the other hand by inserting the traces of $\mathcal L_{s}$ and $\mathcal L_{s+1}$ from (\ref{trLn2}) we get,
\begin{equation}\label{T-m}
\left\lbrace-(-1)^{n}tr \mathcal L_{s+1}^n+tr \mathcal L_{s}^n\right\rbrace=\sum_{x\in Fix_+T^n}\prod_{k=0}^{n-1}\left( T^{k}(x)\right)^{2s}
\end{equation}
which completes the proof.
\end{proof}
Consider now the dynamical system $\mathcal D_1$ defined in (\ref{dy-sy}) which is closely related to the geodesic flow on the upper half plane mod $PSL(2,\mathbb Z)$. The Ruelle zeta function (\ref{Ruelle zeta}) for the dynamical system $\mathcal D_1$ at $z=1$ is reduced to a zeta function $\eta(s)$, given by
\begin{equation}\label{interm1-formula}
\eta(s)=exp(\sum_{n=1}^\infty\dfrac{1}{n}\sum_{(x,\epsilon)\in FixP_{ex}^n}\prod_{k=0}^{n-1}g(P_{ex}^k(x,\epsilon)))
\end{equation}
From (\ref{dy-map-sl}) we have
\begin{equation}\label{epsil-trnsform}
P_{ex}(x,\epsilon)=(Tx,-\epsilon)
\end{equation}
where $T$ is the Gauss map. Because of the way that the parameter $\epsilon$ is transformed in (\ref{epsil-trnsform}), obviously the odd powers of $P_{ex}$ have no fixed points and therefore the summation in (\ref{interm1-formula}) is restricted to the even integers $n\in\mathbb N$,
\begin{equation}\label{z34}
\eta(s)=exp(\sum_{n\ even}\dfrac{1}{n}\sum_{(x,\epsilon)\in FixP_{ex}^n}\prod_{k=0}^{n-1}g(P_{ex}^k(x,\epsilon)))
\end{equation}
On the other hand by iterating (\ref{epsil-trnsform}) we get
\begin{equation}\label{jkl}
P_{ex}^k(x,\epsilon)=(T^kx,(-1)^k\epsilon))
\end{equation}
Therefore from (\ref{weight-fun}) and (\ref{jkl}) we get
\begin{equation}\label{weifun1}
g(P_{ex}^k(x,\epsilon))=(T^kx)^{2s}
\end{equation}
By inserting (\ref{weifun1}) into (\ref{z34}) we have
\begin{equation}\label{z54}
\eta(s)=exp(\sum_{n\ even}\dfrac{1}{n}\sum_{(x,\epsilon)\in FixP_{ex}^n}\prod_{k=0}^{n-1}(T^kx)^{2s})
\end{equation}
Finally we note that for even $n\in\mathbb N$ every pair of fixed points $(x,\pm1)$ of the map $P_{ex}^n$ corresponds to the fixed point $x$ of the map $T^n$. Consequently the summation over $FixP_{ex}^n$ can be replaced by twice the sum over the set $Fix_+T^n$, defined in (\ref{fix+}), that is
\begin{equation}\label{zztt}
\eta(s)=exp(2\sum_{n\ even}\dfrac{1}{n}\sum_{x\in Fix_+T^n}\prod_{k=0}^{n-1}(T^kx)^{2s})
\end{equation}
The next lemma shows, how $\eta(s)$ is related to the transfer operator.
\begin{lem8}
For $Re(s)>\frac{1}{2}$ the following identity holds
\begin{equation}\label{zeta-trans3}
\dfrac{det(1-\mathcal L_{s+1}^2)}{det(1-\mathcal L_{s}^2)}=\eta(s)
\end{equation}
where $\mathcal L_s$ denotes the transfer operator for $PGL(2,\mathbb Z)$ and the determinants are defined in the sense of Grothendieck.
\end{lem8}
\begin{proof}
As in the previous lemma according to Grothendieck's Fredholm determinant for nuclear operators (\ref{def-Gro-det}), we have
\begin{equation}
det(1-\mathcal L_{s+1}^2)=exp(-\sum_{m}\frac{1}{m}tr\mathcal L^{2m}_{s+1}), \ \ \ \ Re (s)>\frac{1}{2}
\end{equation} 
and
\begin{equation}
det(1-\mathcal L_{s}^2)=exp(-\sum_{m}\frac{1}{m}tr\mathcal L^{2m}_{s}), \ \ \ \ Re (s)>\frac{1}{2}
\end{equation}
Then we get
\begin{equation}
\dfrac{det(1-\mathcal L_{s+1}^2)}{det(1-\mathcal L_{s}^2)}=exp(\sum_{m}\frac{1}{m}\left\lbrace-tr \mathcal L_{s+1}^{2m}+tr \mathcal L_{s}^{2m}\right\rbrace), \ \ \ \ Re(s)>\frac{1}{2}
\end{equation}
or equivalently
\begin{equation}\label{ddjjkk}
\dfrac{det(1-\mathcal L_{s+1}^2)}{det(1-\mathcal L_{s}^2)}=exp(2\sum_{n\ even}\frac{1}{n}\left\lbrace-tr \mathcal L_{s+1}^{n}+tr \mathcal L_{s}^{n}\right\rbrace), \ \ \ \ Re(s)>\frac{1}{2}
\end{equation}
On the other hand from (\ref{trLn2}), by a simple algebraic calculation we have,
\begin{equation}
\left\lbrace-tr \mathcal L_{s+1}^{n}+tr \mathcal L_{s}^{n}\right\rbrace=\sum_{x\in Fix_+T^n}\prod_{k=0}^{n-1}\left( T^{k}(x)\right)^{2s}\ \ \ \ n\ \text{even}
\end{equation}
By replacing this into (\ref{ddjjkk}), we achieve the desired result. 
\end{proof}
\begin{cor5}
Let $\overset{\sim}{\mathcal L}_s$ denote the transfer operator for $PSL(2,\mathbb Z)$. For $Re(s)>\frac{1}{2}$ the following identity holds
\begin{equation}\label{trans-zeta2}
\dfrac{det(1-\overset{\sim}{\mathcal L}_{s+1})}{det(1-\overset{\sim}{\mathcal L}_s)}=\eta(s)
\end{equation}
\end{cor5}
\begin{proof}
The representation of the Mayer transfer operator for $PSL(2,\mathbb Z)$ given in (\ref{Mayer vect}) immediately leads to 
\begin{equation}
det(1-\overset{\sim}{\mathcal L}_s)=det(1-\mathcal L_s)det(1+\mathcal L_s)=det(1-\mathcal L^2_s)
\end{equation}
This together with the previous lemma give the desired result. 
\end{proof}
\section{Selberg zeta function and transfer operator}
In this section we illustrate the connection between the Selberg zeta function and the Mayer transfer operator which is one of the most important aspects of Mayer's theory. First we recall the definition of the Selberg zeta function. Let $\Gamma$ be a Fuchcian group of the first kind. The Selberg zeta function for $\Gamma$ is defined in the domain $Re(s)>1$ by an absolutely convergent infinite product given by (see \cite{Selberg}) 
\begin{equation}
Z_{\Gamma}(s)=\prod_{k=0}^\infty\prod_{\left\lbrace P\right\rbrace_\Gamma }(1-\mathcal N(P)^{-k-s})
\end{equation}
where $P$ runs over all primitive hyperbolic conjugacy classes of $\Gamma$ and $\mathcal N(P)>1$ denotes the norm of $P$. By definition every hyperbolic element $P$ of the group $\Gamma$ is conjugated by an element from $PSL(2,\mathbb R)$ to a 2 by 2 matrix
\begin{equation}
\left( \begin{array}{cc}
\rho&0\\
0&\rho^{-1}\\
\end{array}
\right)
\end{equation}
with $\rho>1$ which gives the norm of $P$ as $\mathcal N(P)=\rho^2$. 
Using the Selberg trace formula it is proved that \\ (1) $Z_{\Gamma}(s)$ has analytic (meromorphic) continuation to the whole complex $s$-plane \\ (2) $Z_{\Gamma}(s)$ satisfies the functional equation \begin{equation}\label{Sel1}
Z_{\Gamma}(1-s)=\Psi(s)Z_{\Gamma}(s)
\end{equation}
with some known function $\Psi$ \\ (3) the nontrivial zeros of $Z_{\Gamma}(s)$ are related to eigenvalues and resonances of the automorphic Laplacian $A(\Gamma)$ for the group $\Gamma$. 

It is well known that there is a one to one correspondence between primitive hyperbolic conjugacy classes of $\Gamma$ represented by $P$ with norm $\mathcal N(P)$ and the prime closed geodesics $c$ on the Riemann surface $H\setminus\Gamma$(with obvious singularities in some cases) with length $\ell(c)$ such that 
\begin{equation}
\mathcal N(P)=e^{\ell(c)}
\end{equation} 
This fact enables us to define the Selberg zeta function in the following equivalent form, that is
\begin{equation}\label{Selzeta2}
Z_{\Gamma}(s)=\prod_{k=0}^\infty\prod_{\left\lbrace c\right\rbrace _\Gamma}(1-exp(-(s+k)\ell(c)))
\end{equation}
where $\left\lbrace c\right\rbrace _\Gamma$ runs over all prime closed geodesics on $H\setminus\Gamma$. 

In the sequel we need an Euler product for the dynamical function $\eta(s)$ defined in (\ref{zztt}),
\begin{equation}\label{z23}
\eta(s)=exp(\sum_{n\ even}\dfrac{1}{n}\sum_{x\in Fix_+T^{n}}(\prod_{k=0}^{n-1}T^kx)^{2s})
\end{equation}
Following Ruelle \cite{Ru1}, we rewrite $\eta(s)$ as an Euler product. To begin with, we recall that for $x\in [0,1]$, the set 
\begin{equation}
\phi=\left\lbrace x, Tx, \ldots, T^kx, \ldots\right\rbrace 
\end{equation}
is an orbit of the Gauss map $T$ given in (\ref{Gauss-map}). \\The orbit $\phi$ is periodic if there exists an integer $u\in\mathbb N$ such that 
\begin{equation}
T^ux=x
\end{equation}\\
The integer $u\in \mathbb N$ is called the period of the periodic orbit $\phi$.\\
We say that a periodic orbit $\phi$ is primitive of minimal period $m$ if $m$ is the minimum of the set of all periods of $\phi$. We denote the set of all primitive periodic orbits of $T$ of minimal period $m$ by $Per(m)$.

We are going now to replace the sum over $Fix_+T^n$ in (\ref{z23}) by a sum over the primitive periodic orbits $Per(m)$. To this end we first introduce a subset $MFix_+T^m\subset Fix_+T^m$, containing the periodic continued fractions with minimum period $m$, then we have,
\begin{equation}
\sum_{x\in Fix_+T^{n}}(\prod_{k=0}^{n-1}T^kx)^{2s}=\sum_{m\vert n} \ \sum_{x\in MFix_+T^{m}}\prod_{k=0}^{(\frac{n}{m})m-1}(T^kx)^{2s}
\end{equation}
or by noting that $x\in MFix_+T^m$ is of period $m$, we have, 
\begin{equation}
\sum_{x\in Fix_+T^{n}}(\prod_{k=0}^{n-1}T^kx)^{2s}=\sum_{m\vert n} \ \sum_{x\in MFix_+T^{m}}\left[ \prod_{k=0}^{m-1}(T^kx)^{2s}\right] ^{\frac{n}{m}}
\end{equation}
Now by replacing the sum over $MFix_+T^m$ by a sum over $Per(m)$ we get,
\begin{equation}\label{perm}
\sum_{x\in Fix_+T^{n}}(\prod_{k=0}^{n-1}T^kx)^{2s}=\sum_{m\vert n} \ \sum_{\phi\in Per(m)}m\left[ \prod_{k=0}^{m-1}(T^kx_\phi)^{2s}\right] ^{\frac{n}{m}}
\end{equation}
where $x_\phi$ is an arbitrary point of the orbit $\phi\in Per(m)$ and the factor $m$ comes from the fact that a periodic orbit $\phi$ passing a point $x\in MFix_+T^m$, contains also the set of points
\begin{equation}
\left\lbrace T^kx\in MFix_+T^m \vert \ k=0, \ldots, m-1\right\rbrace 
\end{equation}
Inserting the formula (\ref{perm}) into (\ref{z23}) we get,
\begin{equation}
\eta(s)=exp\left( 2\sum_{n\ even}\dfrac{1}{n}\sum_{m\vert n} \ \sum_{\phi\in Per(m)}m\left[ \prod_{k=0}^{m-1}(T^kx_\phi)^{2s}\right] ^{\frac{n}{m}}\right) 
\end{equation}
or by rearranging the summation,
\begin{equation}\label{formula-eta}
\eta(s)=exp\left( 2\sum_{r\ even} \sum_{\phi\in Per(r)}\sum_{q=0}^\infty\dfrac{1}{q}\left[ \prod_{k=0}^{r-1}(T^kx_\phi)^{2s}\right] ^{q}\right) 
\end{equation}
Since 
\begin{equation}
-log(1-w)=\sum_{q=1}^\infty\dfrac{1}{q}w^q
\end{equation}
the formula (\ref{formula-eta}) reduces to
\begin{equation}
\eta(s)=exp\left( -2\sum_{r\ even} \sum_{\phi\in Per(r)}log\left[ 1-\prod_{k=0}^{r-1}(T^kx_\phi)^{2s}\right] \right) 
\end{equation}
or
\begin{equation}
\eta(s)=exp\left( \sum_{r\ even} \sum_{\phi\in Per(r)}log\left[ 1-\prod_{k=0}^{r-1}(T^kx_\phi)^{2s}\right] ^{-2}\right) 
\end{equation}
which leads finally to the desired Euler product,
\begin{equation}\label{E-p}
\eta(s)=\prod_{r\ even} \prod_{\phi\in Per(r)}\dfrac{1}{(1-\prod_{k=0}^{r-1}(T^kx_\phi)^{2s})^{2}}
\end{equation}

The Euler product for $\eta(s)$ given above is crucial for the following lemma which is a bridge between Mayer transfer operator and Selberg zeta function.
\begin{lem9}
For $Re(s)>1$ the following identity holds
\begin{equation}\label{interm-step}
Z(s)^{-1}=\prod_{l=0}^\infty\eta(s+l)
\end{equation}
where $Z(s)$ is the Selberg zeta function for the group $PSL(2,\mathbb Z)$ and $\eta(s)$ is defined in (\ref{z23}) with the Euler product given in (\ref{E-p}). 
\end{lem9}
\begin{proof}
First we need to rewrite (\ref{E-p}) as a product over primitive priodic orbits of the map $P_{ex}$ defined in (\ref{dy-map-sl}). Let $\widehat{Per}(r)$ denotes the set of primitive periodic orbits of minimal period $r=2a$ with $a\in \mathbb N$ for the map $P_{ex}$. According to (\ref{jkl}) an element $\widehat{\phi}\in\widehat{Per}(r)$ passing the point $(x,\epsilon)$ is defined by the following set,
\begin{equation}
\widehat{\phi}=\left\lbrace (T^kx,(-1)^k\epsilon)\ \vert \ k=0, \ldots, r-1, \ \epsilon=\pm1\right\rbrace 
\end{equation}
where $x\in [0,1]$ and $\epsilon=\pm1$. We note that all periodic orbits of $P_{ex}$ have even period $r$. Obviously every two elements of $\widehat{Per}(r)$ corresponds to an element of $Per(r)$ and also noting that the terms in the product (\ref{E-p}) do not depend on $\epsilon$, the power $2$ in the denominator of (\ref{E-p}) disappears if we replace $Per(r)$ by $\widehat{Per}(r)$,
\begin{equation}\label{E-pp}
\eta(s)=\prod_{r\ even} \prod_{\widehat{\phi}\in \widehat{Per}(r)}\dfrac{1}{1-\prod_{k=0}^{r-1}(T^kx_{\widehat{\phi}})^{2s}}
\end{equation} 
where $x_{\widehat{\phi}}$ is an arbitrary point of $\widehat{\phi}$.
On the other hand according to Series \cite{bowen}, Adler and Flatto \cite{adler} there is an one to one correspondence between $\widehat{Per}(r)$ and the set of primitive periodic orbits $\vartheta$ on the unit tangent bundle $T_1M$, $M=PSL(2,\mathbb Z)\backslash \mathbb H$, with the period (see \cite{Mayer92})
\begin{equation}
\tau(\vartheta)=-2ln\prod_{k=0}^{r-1}T^kx_{\widehat{\phi}}
\end{equation}

These facts recover the physical situation behind the abstract number theoretic appearance of the problem, leading to the following equivalent formula for the dynamical zeta function in (\ref{E-pp}),
\begin{equation}
\eta(s)=\prod_{{\left\lbrace \vartheta\right\rbrace }_{\Gamma}} \dfrac{1}{1- exp(-s\tau(\vartheta))}
\end{equation}
where ${\left\lbrace \vartheta\right\rbrace }_{\Gamma}$ denotes the set of all primitive periodic orbits on $T_1M$, $M=\Gamma\backslash \mathbb H$ with $\Gamma=PSL(2,\mathbb Z)$. But there is also an one to one correspondence between primitive periodic orbits on $T_1M$ and closed geodesics $c$ on $M$ with length $\ell(c)=\tau(\vartheta)$. Thus we have another equivalent formula for our zeta function
\begin{equation}\label{step111}
\eta(s)=\prod_{{\left\lbrace c\right\rbrace }_{\Gamma}} \dfrac{1}{1- exp(-\beta \ell(c))}
\end{equation}
Here ${\left\lbrace c\right\rbrace }_{\Gamma}$ denotes the set of primitive closed geodesics on $M=\Gamma\backslash \mathbb H$ with $\Gamma=PSL(2,\mathbb Z)$ and $\ell(c)$ is the length of the closed geodesic $c$. Note that because of the unity of the tangent bundle,    $\ell(c)=\tau(\vartheta)$.
Finally inserting (\ref{step111}) into (\ref{Selzeta2}) provides the desired result.
\end{proof}
This lemma immediately leads to the most important feature of the Mayer transfer operator theory namely,
\begin{th1}\label{th1}
For $Res>1$ the following identities holds
\begin{equation}\label{www}
det(1-\overset{\sim}{\mathcal L}_s)=Z(s)
\end{equation}
and
\begin{equation}\label{wwww}
det(1-\mathcal L_s^2)=Z(s)
\end{equation}
where $Z(s)$ denotes the Selberg zeta function for the group $PSL(2,\mathbb Z)$ and $\overset{\sim}{\mathcal L}_s$ is the transfer operator also for $PSL(2,\mathbb Z)$ and $\mathcal L_\beta$ is the transfer operator for $PGL(2,\mathbb Z)$. 
\end{th1}
\begin{proof}
It is enough to insert (\ref{trans-zeta2}) and (\ref{zeta-trans3}) into (\ref{interm-step}).
\end{proof}
\begin{rem3}
The domain of validity of (\ref{www}) and (\ref{wwww}) extends immediately to $s\in\mathbb C$ except some possible small singular set, since the Selberg zeta function is a meromorphic function on the whole plane $\mathbb C$.
\end{rem3}
\section{Number theoretic approach to connection of Selberg zeta function and Mayer transfer operator}
In the previous section based on the one to one correspondence between the primitive periodic orbits of $P_{ex}$ and primitive periodic orbits on the phase space $T_1M$, we passed from the number theoretic appearance of the problem to its dynamical nature. In this physical realization of the problem we could see the connection of the Selberg zeta function and the Mayer transfer operator.

Efrat \cite{Efrat} and later, Lewis and Zagier \cite{LZ} reproved Mayer's result in a purely number theoretic approach. In this section we are going to illustrate the alternative approach of Lewis and Zagier. First we introduce their notations. Let $\gamma\in GL(2,\mathbb Z)$ acts on $D_r$ via linear fractional transformation. The right action of the semigroup 
\begin{equation}\label{xi}
\Xi=\left\lbrace \gamma=\left( \begin{array}{cc}
a&b\\
c&d\\
\end{array}
\right)\in GL(2,\mathbb Z) \ \vert \ \gamma(D_r) \subseteq D_r\right\rbrace 
\end{equation}
on the space $B(D_r)$ is given by
\begin{equation}
\pi_s\left( \begin{array}{cc}
a&b\\
c&d\\
\end{array}
\right)f(z)=(cz+d)^{-2s}f(\dfrac{az+b}{cz+d}) 
\end{equation}
where $D_r$ and $B(D_r)$ are the same as in section.2.0.
Then the Mayer transfer operator for $PGL(2,\mathbb Z)$ in the domain $Re(s)>\frac{1}{2}$ can be represented by 
\begin{equation}
\mathcal L_s=\sum_{n=1}^\infty \pi_s\left( \begin{array}{cc}
0&1\\
1&n\\
\end{array}
\right)
\end{equation}
We note that $\left( \begin{array}{cc}
0&1\\
1&n\\
\end{array}
\right)\in \Xi$ for all $n\in \mathbb N$. \\The set of so called reduced elements of $SL(2,\mathbb Z)$ is defined by 
\begin{equation}
\text{Red}=\left\lbrace \left( \begin{array}{cc}
a&b\\
c&d\\
\end{array}
\right)\in SL(2,\mathbb Z) \ \vert \ 0\leq a\leq b, \ c\leq d\right\rbrace 
\end{equation}
For a hyperbolic element $\gamma$ of $SL(2,\mathbb Z)$ a positive integer $k=k(\gamma)$ is defined to be the largest integer such that $\gamma=\gamma_1^k$ for some $\gamma_1\in SL(2,\mathbb Z)$. Therefore for a primitive hyperbolic element we have $k=1$.
Now we quote the heart of the proof of Lewis and Zagier, based on a classical reduction theory for quadratic forms, in the form of a lemma whose proof one can find in \cite{LZ}.
\begin{lem10}\label{lem10}
\item[\textbf{1)}] Every reduced matrix $\gamma\in Red$ has a unique decomposition as the following product
\begin{equation}
\left( \begin{array}{cc}
0&1\\
1&n_1\\
\end{array}
\right)\ldots \left( \begin{array}{cc}
0&1\\
1&n_{2l}\\
\end{array}
\right) \ \ \ \ \ n_1, \ldots,n_{2l}\geq 1
\end{equation}
for a unique positive integer $l=l(\gamma)$, called the length of $\gamma$. 
\item[\textbf{2)}] There are $2l(\gamma)/k(\gamma)$ reduced representatives with the same length $l(\gamma)$ in every hyperbolic conjugacy class of $SL(2,\mathbb Z)$containing $\gamma$.
\end{lem10}
Next consider the Selberg zeta function for the group $\Gamma=SL(2,\mathbb Z)$ for $Re(s)>1$, 
\begin{equation}\label{SelSel}
Z_{\Gamma}(s)=\prod_{m=0}^\infty\prod_{\left\lbrace P\right\rbrace_\Gamma }(1-\mathcal N(P)^{-m-s})
\end{equation}
By taking the logarithm of both sides and replacing the Taylor expansion of $\text{log}(1-\mathcal N(P)^{-m-s})$ we get
\begin{equation} 
-\text{log} Z(s)=\sum_{\left\lbrace P\right\rbrace_\Gamma}\sum_{m=0}^\infty \sum_{k=1}^\infty \dfrac{1}{k}\mathcal N(P)^{-k(m+s)}
\end{equation}
The absolute convergence of the product (\ref{SelSel}) leads to absolute convergence of the sum above. Thus the interchange of the sums over $m$ and $k$ is allowed,
\begin{equation}
-\text{log} Z(s)=\sum_{\left\lbrace P\right\rbrace_\Gamma}\sum_{k=1}^\infty \dfrac{1}{k}\mathcal N(P)^{-ks}\sum_{m=0}^\infty \mathcal N(P)^{-km}
\end{equation}
but the sum over m is the Taylor expansion of $(1-\mathcal N(P)^{-k})^{-1}$, thus
\begin{equation}
-\text{log} Z(s)=\sum_{\left\lbrace P\right\rbrace_\Gamma} \sum_{k=1}^\infty \dfrac{1}{k}\dfrac{\mathcal N(P)^{-ks}}{1-\mathcal N(P)^{-k}}
\end{equation}
Next using the fact that $\mathcal N(P)^k=\mathcal N(P^k)$ we can consider the double sum over  $k$ and $\left\lbrace P\right\rbrace_\Gamma$, the primitive hyperbolic conjugacy classes, as a single sum over all, not only the primitive hyperbolic conjugacy classes, denoted by $\left\lbrace \gamma\right\rbrace_\Gamma$,
\begin{equation}
-\text{log} Z(s)=\sum_{\left\lbrace \gamma\right\rbrace_\Gamma} \dfrac{1}{k(\gamma)}\dfrac{\mathcal N(\gamma)^{-s}}{1-\mathcal N(\gamma)^{-1}}
\end{equation}
The second part of Lemma.\ref{lem10}  enables us to replace the sum over the hyperbolic conjugacy classes by the the sum over the set of reduced matrices Red,
\begin{equation}\label{interff}
-\text{log} Z(s)=\sum_{\gamma\in \text{Red}} \dfrac{1}{2l(\gamma)}\dfrac{\mathcal N(\gamma)^{-s}}{1-\mathcal N(\gamma)^{-1}}
\end{equation} 
Now we need the following lemma,
\begin{lem11}
The trace of the operator $\pi_s(\gamma)$ acting on $B(D_r)$ for $\gamma\in \Xi$, is given by  \begin{equation}\label{trtr1}
tr(\pi_s(\gamma))=\dfrac{\mathcal N(\gamma)^{-s}}{1-\mathcal N(\gamma)^{-1}}
\end{equation}
\end{lem11}
\begin{proof}
Let $\psi_\gamma$ denotes the action of $\gamma\in\Xi$ on $\overline{D}_r$,
\begin{equation}
\psi_\gamma(z)=\gamma z:=\dfrac{az+b}{cz+d}, \ \ \ \gamma=\left(\begin{array}{cc}
a&b\\
c&d\\
\end{array}
\right)
\end{equation}
We also put,
\begin{equation}
j(\gamma,z)=cz+d
\end{equation}
Then the operator $\pi_s(\gamma)$ is written in the following form,
\begin{equation}
\pi_s(\gamma)f(z)=j(\gamma,z)^{-2s}f(\psi_\gamma(z))
\end{equation}
But according to the definiton of $\Xi$ in (\ref{xi}), $\psi_\gamma$ maps $\overline{D}_r$ strictly inside itself, thus we can apply lemma.8 to get the eigenvalues of $\pi_s(\gamma)$ which are all simple. These eigenvalues are given by,
\begin{equation}\label{sorme}
\lambda_m(\gamma)=j(\gamma,x^*)^{-2s}\left[\dfrac{d\psi_\gamma}{dz}\vert_{z=x^*}\right]^m
\end{equation}
where $x^*$ is the unique fixed point of $\psi_\gamma$ in $\overline{D}_r$. The existence of a unique fixed point comes from remark.2.
On the other hand we note that,
\begin{equation}
\dfrac{d\psi_\gamma}{dz}=\dfrac{1}{j(\gamma,z)^2}
\end{equation}
thus (\ref{sorme}) reduces to
\begin{equation}
\lambda_m(\gamma)=j(\gamma,x^*)^{-2s-2m}
\end{equation}
The sum of all $\lambda_m(\gamma)$ gives the trace of $\pi_s(\gamma)$,
\begin{equation}\label{klkl23}
tr(\pi_s(\gamma))=\sum_{m=1}^\infty j(\gamma,x^*)^{-2s-2m}=\dfrac{j(\gamma,x^*)^{-2s}}{1-j(\gamma,x^*)^{-2}}
\end{equation}
To complete the proof we must show that $\mathcal N(\gamma)^{-1}=j(\gamma,x^*)^{-2}$. To this end we note that,
\begin{equation}
j(g\gamma g^{-1},gx^*)=j(\gamma,x^*), \ \ \ \ g\in SL(2,\mathbb R)
\end{equation}
but there exists a $g\in SL(2,\mathbb R)$ such that
\begin{equation}
gx^*=0, \ \ \ \ \ g\gamma g^{-1}=\left(\begin{array}{cc}
\rho^{-1}&0\\
0&\rho\\
\end{array}
\right)\in \Xi \ \ \ \ \rho>1
\end{equation}
Thus
\begin{equation}\label{lemtrik}
j(\gamma,x^*)=j(g\gamma g^{-1},gx^*)=\rho=\sqrt{\mathcal N(\gamma)}
\end{equation}
where the last identity comes from the definition of norm in section.6. Inserting (\ref{lemtrik}) in to (\ref{klkl23}) completes the proof.
\end{proof}
We note that $\text{Red}\subset\Xi$, thus we can insert (\ref{trtr1}) in (\ref{interff}),
\begin{equation}
-\text{log} Z(s)=tr\left( \sum_{\gamma\in \text{Red}} \dfrac{1}{2l(\gamma)}\pi_s(\gamma)\right) 
\end{equation} 
but according to part 1) of the Lemma.\ref{lem10} we can write
\begin{equation}
-\text{log} Z(s)=tr\left( \sum_{l=1}^\infty\dfrac{1}{2l}\left( \sum_{n=1}^\infty\pi_s\left( \begin{array}{cc}
0&1\\
1&n\\
\end{array}
\right)\right)^{2l} \right) 
\end{equation} 
That is, we have
\begin{equation}
-\text{log} Z(s)=\sum_{l=1}^\infty\dfrac{1}{2l}tr\left( \mathcal L_s\right)^{2l} 
\end{equation} 
or 
\begin{equation}
Z(s)=exp\left( -\sum_{l=1}^\infty\dfrac{1}{2l}tr\left( \mathcal L_s\right)^{2l}\right)  
\end{equation} 
Finally the right hand side of the equation above is nothing than the Fredholm determinant of $\mathcal L_s^2$ thus we get the desired result, namely
\begin{equation}
Z(s)=det(1-\mathcal L_s^2)
\end{equation} 

\textbf{Acknowledgement.} We would like to thank Dieter Mayer for several important remarks and we would like to say also that all possible mistakes in the text belong to us but not to Mayer's theory we presented in this paper.\\
In this work the first author was supported by DAAD and the International Center of TU Clausthal.

\end{document}